\documentclass[fleqn,10pt]{wlscirep}
\usepackage[utf8]{inputenc}
\usepackage[T1]{fontenc}
\usepackage{lineno}
\usepackage{bbding}
\usepackage{multirow}

\newcommand{\livia}[1]{\textcolor{black}{{#1}}}
\title{High-resolution segmentations of the hypothalamus and its subregions for training of segmentation models}

\author[1,2]{Livia Rodrigues \Envelope}
\author[3,4]{Martina Bocchetta}
\author[1]{Oula Puonti}
\author[1]{Douglas Greve}
\author[5]{Ana Carolina Londe}
\author[5]{Marcondes França}
\author[5]{Simone Appenzeller}
\author[2]{Leticia Rittner\ddag}
\author[1,6,7]{Juan Eugenio Iglesias\ddag}

\affil[1]{Massachusetts General Hospital, Harvard Medical School}
\affil[2]{Universidade Estadual de Campinas, School of Electrical and Computer Engineering}
\affil[3]{Dementia Research Centre, Department of Neurodegenerative Disease, UCL Queen Square Institute of Neurology, University College London, London, United Kingdom}
\affil[4]{Centre for Cognitive and Clinical Neuroscience, Division of Psychology, Department of Life Sciences, College of Health, Medicine and Life Sciences, Brunel University London, United Kingdom}
\affil[5]{Universidade Estadual de Campinas - School of Medical Sciences}
\affil[6]{Centre for Medical Image Computing, University College London}
\affil[7]{Computer Science and Artificial Intelligence Laboratory,  Massachusetts Institute of Technology}

\affil[$\dag$]{Equal Contribution}
\affil[\Envelope]{l180545@dac.unicamp.br}

\begin{abstract}
\footnotetext{ \ddag Equal contribution}
Segmentation of brain structures on magnetic resonance imaging (MRI) is a highly relevant neuroimaging topic, as it is a prerequisite for different analyses such as volumetry or shape analysis.  Automated segmentation facilitates the study of brain structures in larger cohorts when compared with manual segmentation, which is time-consuming. However, the development of most automated methods relies on large and manually annotated datasets, which limits the generalizability of these methods. Recently, new techniques using synthetic images have emerged, reducing the need for manual annotation. 
Here we provide \livia{HELM, \textbf{H}ypothalamic \textit{\textbf{e}x vivo} \textbf{L}abel \textbf{M}aps}, a dataset composed of label maps built from publicly available ultra-high resolution~\textit{ex vivo} MRI from 10 whole hemispheres, which can be used to develop segmentation methods using synthetic data. The label maps are obtained with a combination of manual labels for the hypothalamic regions and automated segmentations for the rest of the brain, and mirrored to simulate entire brains. We also provide the pre-processed~\textit{ex vivo} scans, as this dataset can support future projects to include other structures after these are manually segmented.

\end{abstract}
\begin{document}

\flushbottom
\maketitle

\thispagestyle{empty}

\section*{Background \& Summary}

The hypothalamus is a small cone-shaped region located in the medial part of the brain, comprising small subnuclei containing the cell bodies of various neuron subtypes. It plays a significant role in maintaining body homeostasis and regulating sleep, body temperature, appetite, and emotions
~\cite{saper2014hypothalamus, neudorfer2020high, vercruysse2018hypothalamic}. Many authors have already linked hypothalamic volume variation with
numerous conditions and diseases such as Alzheimer’s disease~\cite{piyush2014analysis}, Huntington's disease~\cite{gabery2015volumetric, bartlett2019investigating}, Behavioral Variant Frontotemporal Dementia (bvFTD) ~\cite{bocchetta2015detailed, piguet}, Amyotrophic Lateral Sclerosis (ALS)~\cite{gorges2017hypothalamic, ahmed}, among others~\cite{seong2019hypothalamic, modi2019individual, wolfe2015focal, gutierrez1998hypothalamic}. However, to be able to measure volume differences in the hypothalamus, all these neuroimaging studies require the delineation of hypothalamic boundaries. Despite being considered the gold standard, manual annotation is a costly and time-consuming task, requiring neuroanatomical expertise and dedicated resources. When applied to the hypothalamus, the challenges are further compounded by its small size and the limited contrast it exhibits with neighboring tissues. This difficulty poses an obstacle to the development of large-scale studies on this brain structure.

Large-scale brain imaging studies using magnetic resonance imaging (MRI) have large potential to enhance our understanding of the human brain in both health and disease conditions. However, these studies are often limited by the need for manual annotation. Automated segmentation methods have been developed to circumvent this problem and allow for a greater quantity of data to be utilized. Existing methods include classical atlas-based approaches~\cite{iglesias2015multi, cabezas2011review} and more modern  deep learning networks~\cite{wang2022medical, lopez2020medical}. While deep learning showcases outstanding performance in segmentation problems, these methods only work on a specific type of MRI modality and resolution. For instance, we can find a few automated segmentation models for the hypothalamus focusing on T1w~\cite{rodrigues2022benchmark, billot2020automated, greve2021deep},  however only one is able to segment both T1w and T2w images~\cite{estrada2023fastsurfer}. 
So far, models trained solely with T1-weighted MR images do not perform effectively on other MRI modalities (T2w, PD, FLAIR, and others) due to the so-called “domain gap”.

While domain adaptation techniques~\cite{wang2018deep} can sometimes mitigate this problem, it does not fully close the domain gap.
More recently, synthetic image-based methods have been proposed to obtain networks that generalize well across different dataset~\cite{neff2017generative, chang2020synthetic, iglesias2023ready, billot2023synthseg, iglesias2021joint}. 
Recently, Billot \textit{et al} proposed Synthseg\cite{billot2023synthseg},  a new segmentation method trained solely with synthetic images derived from label maps. By randomizing the appearance and resolution of the synthetic scans continuously during training, SynthSeg can readily segment images of any contrast and resolution during testing, without retraining

Therefore, in this article, we introduce a dataset of 3D label maps that may be used to train deep-learning-based networks using synthetic data. The presented dataset is derived from 10 \textit{post mortem} ultra-high MRI acquisitions of brains provided by the Distributed Archives for Neurophysiology Data Integration (DANDI Archive)~\cite{dandi}.
Using this sample of~\textit{post mortem} MRI as a starting point, we trained a deep learning network capable of segmenting the hypothalamus and its subregions across various MRI modalities and resolutions. 

The label maps used for the network training are openly available for further research. Also, we provide the pre-processed~\textit{ex vivo} scans, as this dataset has the potential to be extended to different brain structures by manually segmenting them
\section*{Methods}

\subsection*{Ultra-high resolution~\textit{ex vivo} MRI}

\livia{HELM} is derived from 10~\textit{post mortem} ultra-high MRI acquisitions of brains from the public DANDI Archive~\cite{dandi}. The~\textit{post mortem} images are openly available\footnote{\url{https://dandiarchive.org/dandiset/000026/draft/files?location=}} and comply with all relevant ethical regulations~\cite{costantini2023cellular}. 
\livia{The MRI of the~\textit{ex vivo} brain hemispheres was obtained using multi-echo fast low-angle shot (ME FLASH or MEF) on a 7 T Siemens MR scanner with Repetition Time (TR) of 34 ms, time to echo (TE) of 5.65, 11.95, 18.25, and 24.55 ms, and field of view (FOV) of 192 mm by 81.3 mm. Before the MRI, the specimens were fixed in 10\% formalin for a minimum of 90 days and packed in a 2\% buffered paraformaldehyde solution. The images present an equal distribution of 5 male and 5 female control specimens of individuals who died of natural causes with no clinical diagnoses or neuropathology. As the images come from~\textit{post mortem} brains, the acquisition is free from motion artifacts and has a high resolution, ranging from 120 to 150$\mu$m isotropic, which improves the visualization of hypothalamic boundaries. Further details on the specimens and MRI acquisition can be found in the original publication~\cite{costantini2023cellular}.}
The age at the time of death ranges from 54 to 79 years, with an average of 66.4 $\pm$ 8.46 years and they include only a single hemisphere of the brain \livia{(four right, six left)}. 
\livia{In practice, we left-right flip the right hemispheres and work with 10 left hemispheres. This is common practice when working with \emph{ex vivo} datasets with high resolution but limited sample sizes, e.g., ~\cite{iglesias2015computational, iglesias2018probabilistic}.  We note that combining left and (flipped) right hemispheres into a single model has negligible impact on the subsequent training of machine learning models, due to the strong lateral symmetry of the hypothalamus~\cite{kiss2020functional} and to the aggressive geometric augmentation that is often used in training. 
}

Using these MRI images as a starting point, we performed data pre-processing and generated automated whole-brain segmentation using unsupervised clustering. \livia{Unlike the hypothalamic labels, these automated segmentations of the rest of the brain are used only for image synthesis purposes and not as segmentation targets. Therefore,  they can be noisy and not correspond directly to brain structures -- so unsupervised clustering suffices.}

The focus is on capturing context around the hypothalamus to produce synthetic intensities. Subsequently, we mirrored the hemispheres to create label maps that would serve as synthetic images for the input of the deep neural network without having to segment scans into hemispheres during testing.

\subsection*{Pre-processing of \emph{ex vivo} scans}
\label{preprod}
\begin{figure} [!ht]  
\begin{center}  
	\includegraphics[width=0.4\columnwidth]{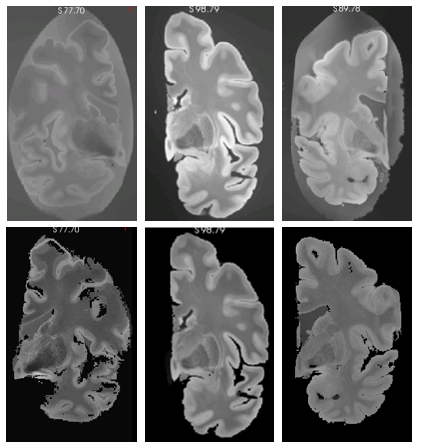}
	\caption{Examples of original \textit{ex vivo} images (up) and images after the pre-processing steps (down). First, we re-orient the images to positive RAS standard and remove non-cerebral elements from the background. Then, we resample the images voxels to $0.3\times0.3\times0.3mm$ isotropic and perform bias field correction. \label{fig:orig}
 }   
\end{center}  
\end{figure}

The first step of the method was the pre-processing of the images. The primary objective at this stage is to standardize the dataset and remove any background elements that could interfere with the subsequent steps. (Fig. \ref{fig:orig}).

\begin{itemize}
    \item \textbf{Orientation}:
Given that the primary objective of the dataset is to facilitate the development of automated segmentation methods, it was essential to ensure that all images were uniformly oriented. We decided to use the positive RAS (right-anterior-superior) standard. \livia{While our dataset includes images of both hemispheres, we left-right flipped all right hemispheres and with the idea of training neural networks to process left hemispheres; these networks can be used to analyze right hemispheres by simply flipping the input MRI scan.
}

    \item \textbf{Background segmentation}: The~\textit{ex vivo} brains were packaged in a bag for scanning, which is discernible in the images and undesirable in our model. To address this issue, we employed a Bayesian automated image segmentation approach adaptive to contrast~\cite{puonti2016fast} to create a mask, which was then utilized to eliminate non-cerebral elements that were not related to brain structures.
 
    \item \textbf{Voxel resampling}: The original images have a voxel resolution ranging from 0.1 to 0.15$mm^3$. While this resolution assists in distinguishing structures during manual segmentation, it significantly prolongs image processing, particularly when the ultimate goal is to employ them in deep learning networks. Therefore, we adjusted the voxel resolution to a constant resolution equal to $0.3\times0.3\times0.3mm$ isotropic, which provides a compromise between a high level of details and being storage- and processing-friendly 

    \item \textbf{Bias Field Correction}:
    Finally, the last step in image pre-processing is the bias field correction~\cite{puonti2016fast}. This step is essential, as in the generation of the whole brain segmentation, we utilize an unsupervised clustering method that can be directly affected by the bias field.
\end{itemize}

\subsection*{Segmentation}
\label{sec:segm}

In the second stage, we generated the label maps for the hypothalamus, its subregions, and the whole brain:

\begin{itemize}
   
    \item \textbf{Hypothalamus manual segmentation:} 
 All images were traced by \livia{one} non-specialist trained by MFJ. For manual segmentation, we relied on protocols focused on~\textit{in vivo} images as described in the literature. In particular, we followed the criteria described in Rodrigues~\textit{et al}~\cite{rodrigues2022benchmark} (whole hypothalamus) and Bocchetta~\textit{et al}~\cite{bocchetta2015detailed} (for the subregions). At 0.3$mm$ isotropic resolution, approximately 40-50 coronal slices include the hypothalamus.

\begin{itemize}
    \item \textbf{Whole structure:}
The segmentation of the whole hypothalamus occurs on coronal view. To ensure the correct delineation of the landmarks, we also simultaneously inspected the sagittal and axial views. In~\textit{in vivo} images, the hypothalamus lies around the third ventricle. However, In the case of the~\textit{ex vivo} images, is not always possible to distinguish the third ventricle, since we only have one hemisphere of the brain. Therefore, on the coronal view, we use the recess dorsal to the hypothalamus to define its most superior boundary (Fig~\ref{fig:exv1}).

\begin{figure} [!ht]  
\begin{center}  
	\includegraphics[width=0.35\columnwidth]{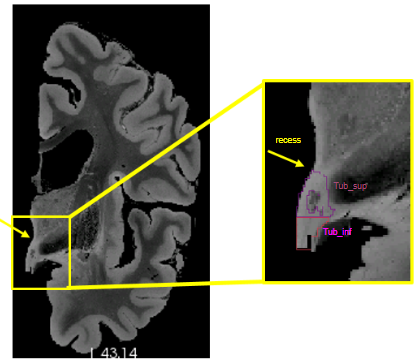}
	\caption{\livia{Recess of the hypothalamus used for the delineation of superior boundary. Tub\char`_sup = Tuberal Superior subregion;Tub\char`_inf = Tuberal inferior subregion\label{fig:exv1}}   }
\end{center}  
\end{figure}

Ventrally, the hypothalamus is defined by the optic tract and the hypothalamic sulcus in the most rostral slices. The most anterior coronal slice is defined as the one where the anterior commissure is visible, while the most caudal coronal slice is where the mammillary bodies (MB) are no longer visible. The mammillary bodies were included in the segmentation, while the fornix and optical tract were excluded from the segmentation.

    \item \textbf{Subregions:} We subdivided the hypothalamus into 5 subregions: Anterior inferior, Anterior Superior, Tuberal inferior, Tuberal Superior, and Posterior. 
    \livia{The manual segmentation of the whole hypothalamus (its outer boundaries with the surrounding structures) relied completely on image contrast and anatomical landmarks. However, the division of the internal subregions relied on a combination of image intensities and geometric criteria, specifically for the subdivision of their rostro–caudal and dorso–ventral borders. Specifically, these geometric criteria included horizontal lines relying on anatomical landmarks, which have been defined and validated
 in Bocchetta~\textit{et al}~\cite{bocchetta2015detailed} (see Table~\ref{tab:landmarks}).} As we only have one hemisphere for each brain, we could not segment both hypothalami for each subject. The most rostral slice of the anterior subregion coincides with the most rostral slice of the hypothalamus when looking at the coronal plane. On~\textit{in vivo} images, the anterior subregion is included from the most rostral coronal slice of the hypothalamus to the most rostral part of the infundibulum. However, for~\textit{ex vivo} images we used the anterior commissure visible from the sagittal view as a landmark to delineate the most caudal part of the anterior regions (Fig.~\ref{fig:exv2}(a)).  The tuberal subregions begin posteriorly to the coronal slice where the anterior regions are visible (as defined by the anterior commissure sagitally) and extend to the most rostral slice where the MB are visible, which are included in the posterior subregion (Fig.~\ref{fig:exv2}(b)). To delineate the superior and inferior portions of both the anterior and tuberal subregions, we drew a horizontal line on the coronal slice connecting the most medial to the most lateral point of the hypothalamus  \livia{(Fig.~\ref{fig:exv2}(c,d))}. 

\begin{table}[!ht]
\begin{center}
\caption{\label{tab:landmarks} \livia{Landmarks used for hypothalamus division into subregions}}
\resizebox{0.80\textwidth}{!}{
\begin{tabular}{c|c|c}
{}& {\textbf{Anterior}}& {\textbf{Tuberal}}\\
\hline
{\textbf{Most rostral} }& {Most rostral coronal slice where}& {Posterior to last coronal slice where the anterior }\\
{\textbf{landmarks}} & {the hypothalamus is visible} & {subregions are visible (defined by the}\\
{}&{}&{anterior commissure on the sagittal view (Fig.~\ref{fig:exv2}(a,b)))}\\
\hline
{\textbf{Most caudal}}& {Defined by the anterior}& {First coronal slice just rostral to the slice where}\\
{\textbf{landmarks}}&{ commissure on the sagittal view }&{the mammillary bodies are clearly visible}\\
\hline
{\textbf{Superior/Inferior}}& \multicolumn{2}{|c}{Defined on all coronal slices by a horizontal line connecting the most}\\
{\textbf{landmarks}} &\multicolumn{2}{|c}{ medial to the most lateral point of the hypothalamus (Fig.~\ref{fig:exv2}(c,d))}\\
\hline
{\textbf{Lateral/medial}}& \multicolumn{2}{|c}{Defined as the same boundaries of the }\\
{\textbf{landmarks}} &\multicolumn{2}{|c}{whole hypothalamus}
\end{tabular} 
}
\end{center}
\end{table}

    \begin{figure} [!ht]  
\begin{center}  
	\includegraphics[width=0.4\columnwidth]{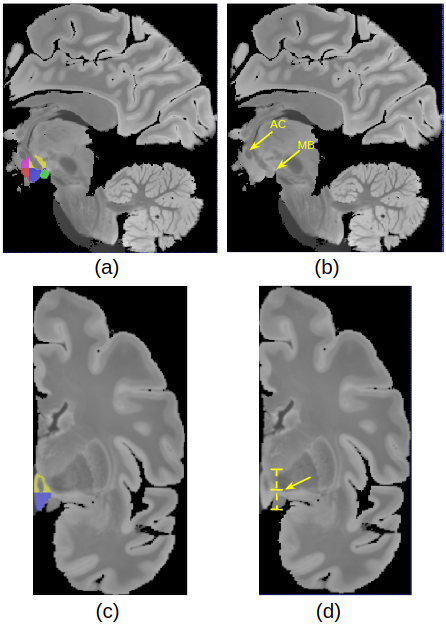}
	\caption{\textbf{Segmentation protocol:} \livia{(a) Subregions delineation: Anterior superior (pink), anterior inferior (red), tuberal superior (yellow), tuberal inferior (blue) and posterior (green) (b)Sagittal landmarks for subregion delineation (AC = Anterior Comissure and MB=mamillary body) (c) Coronal View of superior and inferior tuberal subregions (d) Coronal landmark for superior/inferior division\label{fig:exv2}}   }
\end{center}  
\end{figure}

\end{itemize}

     \item \textbf{Whole brain segmentation}: Unlike the hypothalamic subregions, semantic meaning is not necessary for the labels of the rest of the brain \livia{ -- since they are only used to generate synthetic image contrast, and not as segmentation targets. That means we do not need to correctly delineate the morphological borders of the other structures.} We just need to establish context around the hypothalamus to generate the synthetic images. We decided to use k-means, a non-supervised clustering method, to model the non-hypothalamic tissue based on grayscale levels of the pre-processed images. Seeking to increase data variability, we ran this step for values of k ranging from 4 to 9. As a result, for each of the 10 images, we have 6 different maps, totaling 60 label maps (Fig.~\ref{fig:kmeans})
\begin{figure} [!ht]  
\begin{center}  
	\includegraphics[width=0.45\columnwidth]{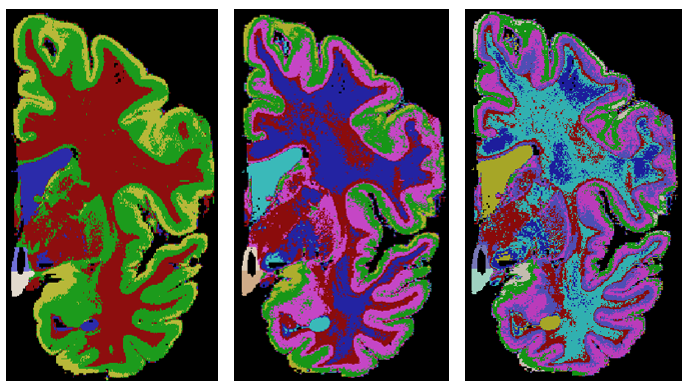}
	\caption{Examples of three different label maps derived from the same image. From left to right: $k=4$, $k=6$, $k=8$ \label{fig:kmeans}}   
\end{center}  
\end{figure}
    
     \item \textbf{Segmentation merge:}
      Finally, it is necessary to merge both segmentations. In this stage, we needed to ensure that there would be no discrepancy during the overlay process. To achieve this, we employed mathematical morphology~\livia{(specifically, a closing operator with a spherical structuring element, radius = 1.2mm) to refine the  delineation of the fornix and  eliminate false positive voxels in the third ventricle area (Fig.~\ref{fig:merge}).}
\end{itemize}

\begin{figure} [!ht]  
\begin{center}  
	\includegraphics[width=0.6\columnwidth]{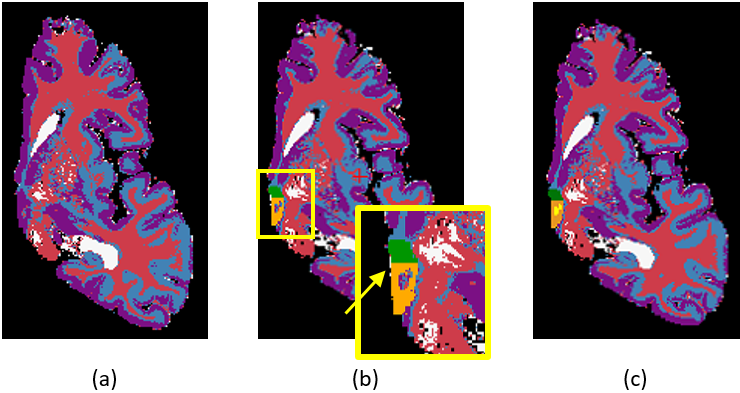}
	\caption{(a) Whole brain segmentation: example with k=4 (b) Manual segmentation simply overlapping the whole brain segmentation. We can see that there are a few inconsistent voxels, that should be labeled either as hypothalamus or background that have different labels. (c) To fix these inconsistencies, a mathematical morphology-based algorithm is applied. \label{fig:merge}}   
\end{center}  
\end{figure}

\subsection*{Hemisphere mirroring}

\begin{figure} [!ht]  
\begin{center}  
	\includegraphics[width=0.65\columnwidth]{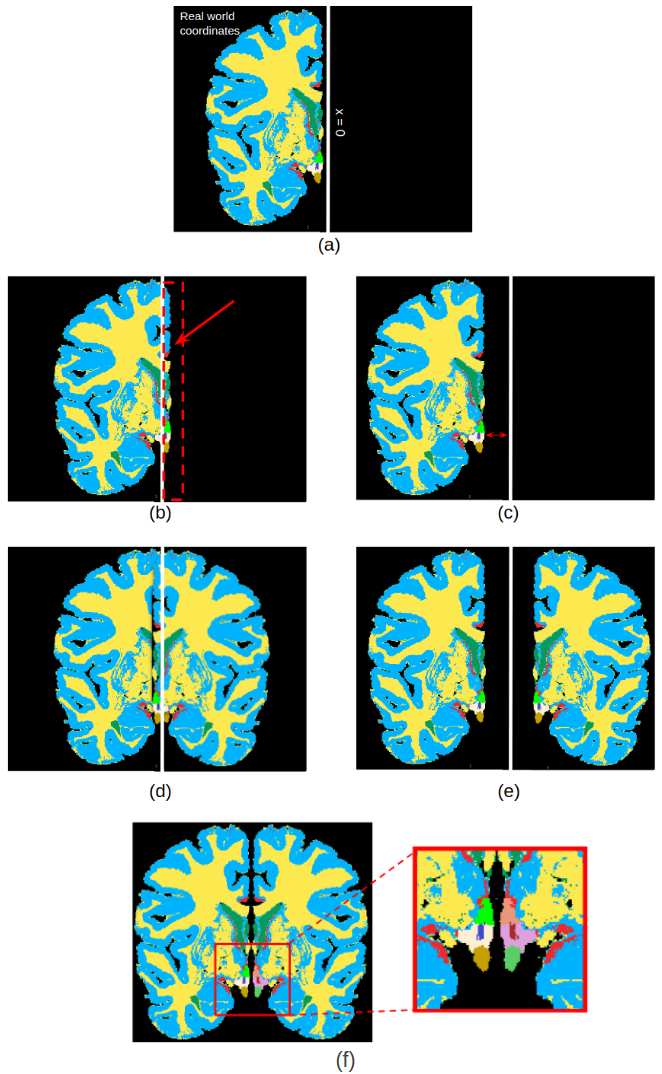}
	\caption{Label map creation: Following the segmentation step, a half-brain label map is generated (a). However, given the hypothalamus's central location within the brain, mirroring is essential to provide contextual information. For the mirroring process, translation and rotation are applied to the RAS coordinates. This involves moving the brain close to the \textit{x=0} axis from the negative side, without surpassing into the positive side. Essentially, a final cost function is computed, penalizing positive values of \textit{x} (b) and high negative values.  Finally, we prevent the overlap between brain hemispheres (d) and also prevent them from ending up at unnaturally distant positions (e). After the optimization and mirroring, we end up with the final label map (f). \label{fig:mirror}}   
\end{center}  
\end{figure}

The~\textit{ex vivo} images were acquired from single brain hemispheres. Given that the hypothalamus is situated in the most medial part of the brain, a model covering a single hemisphere lacks contextual information regarding its surroundings. To address this concern, we mirrored the images to generate a complete brain. In the mirroring process, two major concerns needed to be addressed: the original and mirrored hemispheres should not overlap, and they should be as close as possible to each other (Fig.~\ref{fig:mirror}). 

To tackle this issue, we encode the two constraints into softened versions and combine them into a weighted cost function that is optimized with respect to a rigid transformation T: 

$$\hat{T} = argmin_T  \sum_{i\in\Omega}  \delta[x(v_i;T) > 0]x(v_i;T) - \alpha \sum_{i\in\Omega} x(v_i;T)$$

\noindent where $\Omega$ is the brain mask, $x(v_i; T)$ is the $x$ real-world coordinate after rigid transformation with $T$, $\delta$ is Kronecker's delta, and $\alpha$ a trade-off value that changes according to the image.
\livia{The rationale behind the cost function is the following: assuming a virtual mirror located on the plane $x=0$, we want positive values of $x$ to be penalized (Fig.~\ref{fig:mirror}b) so that the hemisphere does not cross the mirror (i.e., the real and mirrored hemispheres do not overlap, as in Fig.~\ref{fig:mirror}d). This soft constraint corresponds to the first term of the equation. At the same time, the hemisphere should not be too distant from the mirror at $x=0$ (Fig.~\ref{fig:mirror}c), to prevent gaps between the real and mirrored hemispheres(Fig.~\ref{fig:mirror}e). This is encoded into the second term of the equation -- which also includes relative weight $\alpha$ that represents a trade-off between the two terms.}
After optimization, we mirror the transformed hemisphere and merge the real and mirrored hemispheres into one image, ending up with 10 subregions (five from each hemisphere, see Fig.~\ref{fig:mirror}f).

\section*{Data Records}

\livia{HELM is available at \href{https://www.nitrc.org/projects/hsynex_data/}{https://www.nitrc.org/projects/hsynex\char`_data/}(release 0.1)~\cite{hsynex_data} under CC-BY license. The data is in NIfTI format and the image naming system corresponds to the original dataset.}
The following files are provided:

\begin{itemize}
    \item \livia{\textit{preprocessed\char`_exvivo\char`_images.zip}}: \livia{For reproducibility purposes, and to facilitate the development of segmentation methods for other surrounding brain structures, we provide in this file the 10 pre-processed \textit{ex vivo} scans (including reorientation, background segmentation, and bias field correction); see further details in the \textit{Code availability} Section}
   \item \livia{\textit{hypothalamus\char`_manual\char`_seg.zip}}: Manual segmentation of the hypothalamus and its subregions of the~\textit{ex vivo} images following the protocol explained in the Segmentation section. 
   \item  \livia{\textit{whole\char`_brain\char`_segmentation.zip}}: For each subject, there are six whole brain segmentation for context. They are available to be merged with the manual segmentation
   \item \livia{\textit{label\char`_maps\char`_1.zip and label\char`_maps\char`_2.zip}}: Our final label maps using the manual segmentation of the hypothalamus
\end{itemize}

\section*{Technical Validation}
\label{validation}

Finally, we assessed the data quality and usability of the dataset in training neural networks.

\begin{itemize}
    \item \textbf{Manual segmentation quality assessment}: 
      To test the intra-rater reliability of the manual segmentations, one \livia{LMR (trained by MFJ)} manually segmented 5 out of the 10 images a second time, using the same protocol. The images were randomly selected from the total sample and were re-segmented blindly, with the same software and computer settings, after four months from the first manual segmentation. For evaluation, we used the Dice Coefficient and Average Hausdorff Distance~\cite{taha2015metrics},  
      \livia{which are summarized in Table~\ref{tab:inter_tab}. We note that the low image contrast between adjacent hypothalamic subregions and their small size lead to generally low intra-rater Dice. For example, Billot~\textit{et al}~\cite{billot2020automated} reported Dice scores between 0.70 and 0.87 for different subregions of the hypothalamus. Estrada~\textit{et al}~\cite{estrada2023fastsurfer} reported an intra-rater Dice coefficient of 0.82 for the whole hypothalamus. In both studies, the authors highlighted the challenges associated with manual delineation due to the small size and low contrast of the hypothalamus. }
      
\begin{table}[!ht]
\begin{center}
\caption{\label{tab:inter_tab} Intra-rater metrics for 5 subjects.}
\resizebox{0.55\textwidth}{!}{
\begin{tabular}{c|c|c}
{}& {Dice Coefficient}& {Average Hausdorff Distance }\\
\hline
{Whole}& {0.82$\pm$0.03}& {0.37$\pm$0.14}\\
\hline
{Anterior}& {0.65$\pm$0.2}& {0.84$\pm$0.75}\\
\hline
{Tuberal}& {0.78$\pm$0.05}& {0.43$\pm$0.17}\\
\hline
{Posterior}& {0.79$\pm$0.04}& {0.32$\pm$0.08}\\
\end{tabular} 
}
\end{center}
\label{tab:dice}
\end{table}

    \item \textbf{Usability}:
    \livia{The dataset's usability assessment involved implementing a deep learning-based network method for hypothalamus segmentation. We generated synthetic images as Gaussian mixture models (with random means and variances) conditioned on the geometrically augmented labels and blurred them with random kernels to simulate images of different orientations and slice thicknesses~\cite{billot2023synthseg}. We used these images and the corresponding hypothalamic labels to train two cascaded 3D U-Nets~\cite{cicek20163d}: one segmenting the whole hypothalamus, and a second subdividing into subregions. Model validation was conducted using a validation set comprising five different MR sequences (T1, T2, PD, FA, and qT1). The random parameter sampling ensures generalization to any contrast and resolution in the input data~\cite{billot2023synthseg}. For evaluation, we used in vivo MRI from two publicly available datasets: IXI~\cite{IXI} and ADNI~\cite{mueller2005alzheimer}.
    Further information regarding the methodology, such as the datasets utilized for validation and testing, as well as quantitative evaluation metrics, are available in Rodrigues~\textit{et al}.~\cite{rodrigues2024hsynex}}

\end{itemize}

\begin{figure} [!ht]  
\begin{center}  
	\includegraphics[width=0.95\columnwidth]{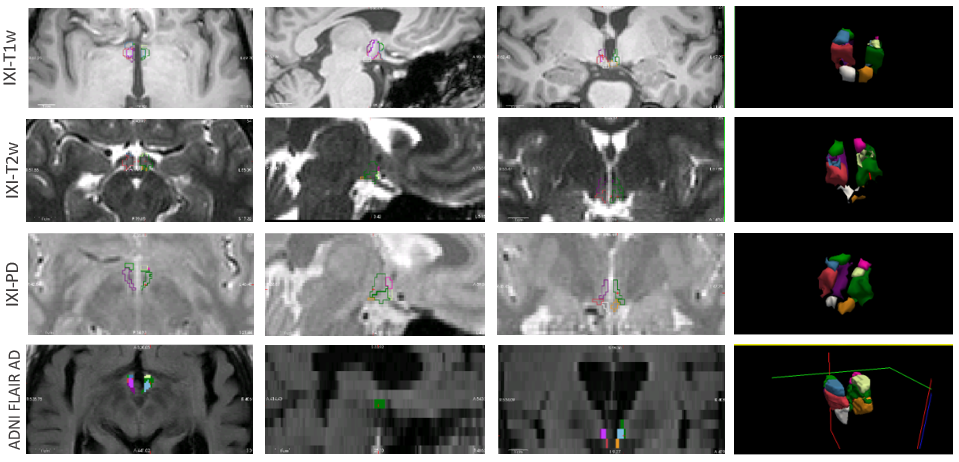}
	\caption{Hypothalamus and subregions segmentation on \textit{in vivo} images using the method trained on the synthetic dataset. The method was capable of segmenting the hypothalamus in T1w, T2w, PD, and FLAIR sequences, the last one presenting a spacing of 5$mm$ \label{fig:qualit}}   
\end{center}  
\end{figure}

\section*{Usage Notes}

The label maps described here could be employed for training different networks dedicated to the hypothalamus or different brain structures. To achieve this, three steps need to be followed:

\begin{itemize}
    \item \textbf{Manual segmentation}: Manually segment the target structure. Given that there are only 10 images, the manual segmentation will not demand as much effort as the ones typically used in supervised learning.
    \item \textbf{Merge with whole brain segmentation}: Merge the manual segmentation with the provided whole brain segmentation. 
    \item \textbf{Mirroring}: Run the provided mirroring codes to generate the final label maps.
\end{itemize}

We also encourage the usage of this dataset in tasks other than segmentation. For instance, we can find in the literature the use of synthetic images applied to registration~\cite{iglesias2023easyreg} and conversion of different MRIs into high-resolution T1 scans~\cite{iglesias2021joint}.

\section*{Code availability}%
 The 10 pre-processed~textit{ex vivo} images, automated brain segmentations, manual hypothalamus segmentations, as well as the final label maps used for training the hypothalamus segmentation model are available on \href{https://www.nitrc.org/projects/hsynex_data/}{https://www.nitrc.org/projects/hsynex\char`_data/}. The codes for creating the label maps are available on \href{https://github.com/liviamarodrigues/hsynex}{https://github.com/liviamarodrigues/hsynex}

\section*{Acknowledgements} 
L.Rodrigues acknowledges the Coordination for the Improvement of Higher Education Personnel (88887.716540/2022-00). M. Bocchetta is supported by a Fellowship award from the Alzheimer’s Society, UK (AS-JF-19a-004-517). J.E.Iglesias acknowledges NIH 1RF1MH123195, 1R01AG070988, 1R01AG070988, 1R01AG070988, and a grant from the Jack Satter foundation. L. Rittner acknowledges CNPq 313598/2020-7 and FAPESP 2013/07559-3. S.Appenzeller acknowledges CAPES Print, CAPES 001 e BRAINN.

\bibliography{cas-refs}

\begin{thebibliography}{10}
\urlstyle{rm}
\expandafter\ifx\csname url\endcsname\relax
  \def\url#1{\texttt{#1}}\fi
\expandafter\ifx\csname urlprefix\endcsname\relax\def\urlprefix{URL }\fi
\expandafter\ifx\csname doiprefix\endcsname\relax\def\doiprefix{DOI: }\fi
\providecommand{\bibinfo}[2]{#2}
\providecommand{\eprint}[2][]{\url{#2}}

\bibitem{saper2014hypothalamus}
\bibinfo{author}{Saper, C.~B.} \& \bibinfo{author}{Lowell, B.~B.}
\newblock \bibinfo{journal}{\bibinfo{title}{The hypothalamus}}.
\newblock {\emph{\JournalTitle{Current Biology}}} \textbf{\bibinfo{volume}{24}}, \bibinfo{pages}{R1111--R1116} (\bibinfo{year}{2014}).

\bibitem{neudorfer2020high}
\bibinfo{author}{Neudorfer, C.} \emph{et~al.}
\newblock \bibinfo{journal}{\bibinfo{title}{A high-resolution in vivo magnetic resonance imaging atlas of the human hypothalamic region}}.
\newblock {\emph{\JournalTitle{Scientific Data}}} \textbf{\bibinfo{volume}{7}}, \bibinfo{pages}{305} (\bibinfo{year}{2020}).

\bibitem{vercruysse2018hypothalamic}
\bibinfo{author}{Vercruysse, P.}, \bibinfo{author}{Vieau, D.} \emph{et~al.}
\newblock \bibinfo{journal}{\bibinfo{title}{Hypothalamic alterations in neurodegenerative diseases and their relation to abnormal energy metabolism}}.
\newblock {\emph{\JournalTitle{Front.{M}ol. {N}eurosci.}}} \textbf{\bibinfo{volume}{11}}, \bibinfo{pages}{2} (\bibinfo{year}{2018}).

\bibitem{piyush2014analysis}
\bibinfo{author}{Piyush, R.} \& \bibinfo{author}{Ramakrishnan, S.}
\newblock \bibinfo{title}{Analysis of sub-anatomic volume changes in {A}lzheimer brain using diffusion tensor imaging}.
\newblock In \emph{\bibinfo{booktitle}{2014 40th Annual Northeast Bioengineering Conference (NEBEC)}}, \bibinfo{pages}{1--2} (\bibinfo{organization}{IEEE}, \bibinfo{year}{2014}).

\bibitem{gabery2015volumetric}
\bibinfo{author}{Gabery, S.}, \bibinfo{author}{Georgiou-Karistianis, N.} \emph{et~al.}
\newblock \bibinfo{journal}{\bibinfo{title}{Volumetric analysis of the hypothalamus in huntington disease using 3{T} {MRI}: The image-hd study}}.
\newblock {\emph{\JournalTitle{PloS one}}} \textbf{\bibinfo{volume}{10}}, \bibinfo{pages}{e0117593} (\bibinfo{year}{2015}).

\bibitem{bartlett2019investigating}
\bibinfo{author}{Bartlett, D.~M.}, \bibinfo{author}{Reyes, A.} \emph{et~al.}
\newblock \bibinfo{journal}{\bibinfo{title}{Investigating the relationships between hypothalamic volume and measures of circadian rhythm and habitual sleep in premanifest huntington's disease}}.
\newblock {\emph{\JournalTitle{Neurobiology of sleep and circadian rhythms}}} \textbf{\bibinfo{volume}{6}}, \bibinfo{pages}{1--8} (\bibinfo{year}{2019}).

\bibitem{bocchetta2015detailed}
\bibinfo{author}{Bocchetta, M.}, \bibinfo{author}{Gordon, E.} \emph{et~al.}
\newblock \bibinfo{journal}{\bibinfo{title}{Detailed volumetric analysis of the hypothalamus in behavioral variant frontotemporal dementia}}.
\newblock {\emph{\JournalTitle{Journal of {N}eurology}}} \textbf{\bibinfo{volume}{262}}, \bibinfo{pages}{2635--2642} (\bibinfo{year}{2015}).

\bibitem{piguet}
\bibinfo{author}{Piguet, O.} \emph{et~al.}
\newblock \bibinfo{journal}{\bibinfo{title}{Eating and hypothalamus changes in behavioral-variant frontotemporal dementia}}.
\newblock {\emph{\JournalTitle{Annals of Neurology}}} \textbf{\bibinfo{volume}{69}}, \bibinfo{pages}{312--319} (\bibinfo{year}{2011}).

\bibitem{gorges2017hypothalamic}
\bibinfo{author}{Gorges, M.}, \bibinfo{author}{Vercruysse, P.} \emph{et~al.}
\newblock \bibinfo{journal}{\bibinfo{title}{Hypothalamic atrophy is related to body mass index and age at onset in amyotrophic lateral sclerosis}}.
\newblock {\emph{\JournalTitle{Journal of {N}eurology, {N}eurosurgery \& {P}sychiatry}}} \textbf{\bibinfo{volume}{88}}, \bibinfo{pages}{1033--1041} (\bibinfo{year}{2017}).

\bibitem{ahmed}
\bibinfo{author}{Ahmed, R.~M.}, \bibinfo{author}{Steyn, F.} \& \bibinfo{author}{Dupuis, L.}
\newblock \bibinfo{journal}{\bibinfo{title}{Hypothalamus and weight loss in amyotrophic lateral sclerosis}}.
\newblock {\emph{\JournalTitle{Handbook of {C}linical {N}eurology}}} \textbf{\bibinfo{volume}{180}}, \bibinfo{pages}{327--338} (\bibinfo{year}{2021}).

\bibitem{seong2019hypothalamic}
\bibinfo{author}{Seong, J.}, \bibinfo{author}{Kang, J.~Y.}, \bibinfo{author}{Sun, J.~S.} \& \bibinfo{author}{Kim, K.~W.}
\newblock \bibinfo{journal}{\bibinfo{title}{Hypothalamic inflammation and obesity: a mechanistic review}}.
\newblock {\emph{\JournalTitle{Archives of pharmacal research}}} \textbf{\bibinfo{volume}{42}}, \bibinfo{pages}{383--392} (\bibinfo{year}{2019}).

\bibitem{modi2019individual}
\bibinfo{author}{Modi, S.}, \bibinfo{author}{Thaploo, D.} \emph{et~al.}
\newblock \bibinfo{journal}{\bibinfo{title}{Individual differences in trait anxiety are associated with gray matter alterations in hypothalamus: Preliminary neuroanatomical evidence}}.
\newblock {\emph{\JournalTitle{Psychiatry Research: Neuroimaging}}} \textbf{\bibinfo{volume}{283}}, \bibinfo{pages}{45--54} (\bibinfo{year}{2019}).

\bibitem{wolfe2015focal}
\bibinfo{author}{Wolfe, F.~H.}, \bibinfo{author}{Auzifas, G.} \emph{et~al.}
\newblock \bibinfo{journal}{\bibinfo{title}{Focal atrophy of the hypothalamus associated with third ventricle enlargement in autism spectrum disorder}}.
\newblock {\emph{\JournalTitle{Neuroreport}}} \textbf{\bibinfo{volume}{26}}, \bibinfo{pages}{1017--1022} (\bibinfo{year}{2015}).

\bibitem{gutierrez1998hypothalamic}
\bibinfo{author}{Gutierrez, M.}, \bibinfo{author}{Garcia, M.}, \bibinfo{author}{Rodriguez, J.}, \bibinfo{author}{Rivero, S.} \& \bibinfo{author}{Jacobelli, S.}
\newblock \bibinfo{journal}{\bibinfo{title}{Hypothalamic-pituitary-adrenal axis function and prolactin secretion in systemic lupus erythematosus}}.
\newblock {\emph{\JournalTitle{Lupus}}} \textbf{\bibinfo{volume}{7}}, \bibinfo{pages}{404--408} (\bibinfo{year}{1998}).

\bibitem{iglesias2015multi}
\bibinfo{author}{Iglesias, J.~E.} \& \bibinfo{author}{Sabuncu, M.~R.}
\newblock \bibinfo{journal}{\bibinfo{title}{Multi-atlas segmentation of biomedical images: a survey}}.
\newblock {\emph{\JournalTitle{Medical image analysis}}} \textbf{\bibinfo{volume}{24}}, \bibinfo{pages}{205--219} (\bibinfo{year}{2015}).

\bibitem{cabezas2011review}
\bibinfo{author}{Cabezas, M.}, \bibinfo{author}{Oliver, A.}, \bibinfo{author}{Llad{\'o}, X.}, \bibinfo{author}{Freixenet, J.} \& \bibinfo{author}{Cuadra, M.~B.}
\newblock \bibinfo{journal}{\bibinfo{title}{A review of atlas-based segmentation for magnetic resonance brain images}}.
\newblock {\emph{\JournalTitle{Computer methods and programs in biomedicine}}} \textbf{\bibinfo{volume}{104}}, \bibinfo{pages}{e158--e177} (\bibinfo{year}{2011}).

\bibitem{wang2022medical}
\bibinfo{author}{Wang, R.} \emph{et~al.}
\newblock \bibinfo{journal}{\bibinfo{title}{Medical image segmentation using deep learning: A survey}}.
\newblock {\emph{\JournalTitle{IET Image Processing}}} \textbf{\bibinfo{volume}{16}}, \bibinfo{pages}{1243--1267} (\bibinfo{year}{2022}).

\bibitem{lopez2020medical}
\bibinfo{author}{L{\'o}pez-Linares~Rom{\'a}n, K.}, \bibinfo{author}{Garc{\'\i}a~Oca{\~n}a, M.~I.}, \bibinfo{author}{Lete~Urzelai, N.}, \bibinfo{author}{Gonz{\'a}lez~Ballester, M.~{\'A}.} \& \bibinfo{author}{Mac{\'\i}a~Oliver, I.}
\newblock \bibinfo{journal}{\bibinfo{title}{Medical image segmentation using deep learning}}.
\newblock {\emph{\JournalTitle{Deep Learning in Healthcare: Paradigms and Applications}}} \bibinfo{pages}{17--31} (\bibinfo{year}{2020}).

\bibitem{rodrigues2022benchmark}
\bibinfo{author}{Rodrigues, L.} \emph{et~al.}
\newblock \bibinfo{journal}{\bibinfo{title}{A benchmark for hypothalamus segmentation on t1-weighted mr images}}.
\newblock {\emph{\JournalTitle{NeuroImage}}} \textbf{\bibinfo{volume}{264}}, \bibinfo{pages}{119741} (\bibinfo{year}{2022}).

\bibitem{billot2020automated}
\bibinfo{author}{Billot, B.}, \bibinfo{author}{Bocchetta, M.} \emph{et~al.}
\newblock \bibinfo{journal}{\bibinfo{title}{Automated segmentation of the hypothalamus and associated subunits in brain {MRI}}}.
\newblock {\emph{\JournalTitle{Neuro{I}mage}}} \textbf{\bibinfo{volume}{223}}, \bibinfo{pages}{117287} (\bibinfo{year}{2020}).

\bibitem{greve2021deep}
\bibinfo{author}{Greve, D.~N.} \emph{et~al.}
\newblock \bibinfo{journal}{\bibinfo{title}{A deep learning toolbox for automatic segmentation of subcortical limbic structures from mri images}}.
\newblock {\emph{\JournalTitle{Neuroimage}}} \textbf{\bibinfo{volume}{244}}, \bibinfo{pages}{118610} (\bibinfo{year}{2021}).

\bibitem{estrada2023fastsurfer}
\bibinfo{author}{Estrada, S.} \emph{et~al.}
\newblock \bibinfo{journal}{\bibinfo{title}{Fastsurfer-hypvinn: Automated sub-segmentation of the hypothalamus and adjacent structures on high-resolutional brain mri}}.
\newblock {\emph{\JournalTitle{arXiv preprint arXiv:2308.12736}}}  (\bibinfo{year}{2023}).

\bibitem{wang2018deep}
\bibinfo{author}{Wang, M.} \& \bibinfo{author}{Deng, W.}
\newblock \bibinfo{journal}{\bibinfo{title}{Deep visual domain adaptation: A survey}}.
\newblock {\emph{\JournalTitle{Neurocomputing}}} \textbf{\bibinfo{volume}{312}}, \bibinfo{pages}{135--153} (\bibinfo{year}{2018}).

\bibitem{neff2017generative}
\bibinfo{author}{Neff, T.}, \bibinfo{author}{Payer, C.}, \bibinfo{author}{Stern, D.} \& \bibinfo{author}{Urschler, M.}
\newblock \bibinfo{title}{Generative adversarial network based synthesis for supervised medical image segmentation}.
\newblock In \emph{\bibinfo{booktitle}{Proc. OAGM and ARW joint Workshop}}, vol.~\bibinfo{volume}{3}, \bibinfo{pages}{4} (\bibinfo{year}{2017}).

\bibitem{chang2020synthetic}
\bibinfo{author}{Chang, Q.} \emph{et~al.}
\newblock \bibinfo{title}{Synthetic learning: Learn from distributed asynchronized discriminator gan without sharing medical image data}.
\newblock In \emph{\bibinfo{booktitle}{Proceedings of the IEEE/CVF conference on computer vision and pattern recognition}}, \bibinfo{pages}{13856--13866} (\bibinfo{year}{2020}).

\bibitem{iglesias2023ready}
\bibinfo{author}{Iglesias, J.~E.}
\newblock \bibinfo{journal}{\bibinfo{title}{A ready-to-use machine learning tool for symmetric multi-modality registration of brain mri}}.
\newblock {\emph{\JournalTitle{Scientific Reports}}} \textbf{\bibinfo{volume}{13}}, \bibinfo{pages}{6657} (\bibinfo{year}{2023}).

\bibitem{billot2023synthseg}
\bibinfo{author}{Billot, B.} \emph{et~al.}
\newblock \bibinfo{journal}{\bibinfo{title}{Synthseg: Segmentation of brain mri scans of any contrast and resolution without retraining}}.
\newblock {\emph{\JournalTitle{Medical image analysis}}} \textbf{\bibinfo{volume}{86}}, \bibinfo{pages}{102789} (\bibinfo{year}{2023}).

\bibitem{iglesias2021joint}
\bibinfo{author}{Iglesias, J.~E.} \emph{et~al.}
\newblock \bibinfo{journal}{\bibinfo{title}{Joint super-resolution and synthesis of 1 mm isotropic mp-rage volumes from clinical mri exams with scans of different orientation, resolution and contrast}}.
\newblock {\emph{\JournalTitle{Neuroimage}}} \textbf{\bibinfo{volume}{237}}, \bibinfo{pages}{118206} (\bibinfo{year}{2021}).

\bibitem{dandi}
\bibinfo{author}{Mazzamuto, G.} \emph{et~al.}
\newblock \bibinfo{journal}{\bibinfo{title}{Human brain cell census for ba 44/45 (version draft).}}
\newblock {\emph{\JournalTitle{DANDI archive}}} \url{https://doi.org/10.80507/dandi.123456/0.123456.1234} (\bibinfo{year}{2004}).

\bibitem{costantini2023cellular}
\bibinfo{author}{Costantini, I.} \emph{et~al.}
\newblock \bibinfo{journal}{\bibinfo{title}{A cellular resolution atlas of broca’s area}}.
\newblock {\emph{\JournalTitle{Science Advances}}} \textbf{\bibinfo{volume}{9}}, \bibinfo{pages}{eadg3844} (\bibinfo{year}{2023}).

\bibitem{iglesias2015computational}
\bibinfo{author}{Iglesias, J.~E.} \emph{et~al.}
\newblock \bibinfo{journal}{\bibinfo{title}{A computational atlas of the hippocampal formation using ex vivo, ultra-high resolution mri: application to adaptive segmentation of in vivo mri}}.
\newblock {\emph{\JournalTitle{Neuroimage}}} \textbf{\bibinfo{volume}{115}}, \bibinfo{pages}{117--137} (\bibinfo{year}{2015}).

\bibitem{iglesias2018probabilistic}
\bibinfo{author}{Iglesias, J.~E.} \emph{et~al.}
\newblock \bibinfo{journal}{\bibinfo{title}{A probabilistic atlas of the human thalamic nuclei combining ex vivo mri and histology}}.
\newblock {\emph{\JournalTitle{Neuroimage}}} \textbf{\bibinfo{volume}{183}}, \bibinfo{pages}{314--326} (\bibinfo{year}{2018}).

\bibitem{kiss2020functional}
\bibinfo{author}{Kiss, D.~S.} \emph{et~al.}
\newblock \bibinfo{journal}{\bibinfo{title}{Functional aspects of hypothalamic asymmetry}}.
\newblock {\emph{\JournalTitle{Brain Sciences}}} \textbf{\bibinfo{volume}{10}}, \bibinfo{pages}{389} (\bibinfo{year}{2020}).

\bibitem{puonti2016fast}
\bibinfo{author}{Puonti, O.}, \bibinfo{author}{Iglesias, J.~E.} \& \bibinfo{author}{Van~Leemput, K.}
\newblock \bibinfo{journal}{\bibinfo{title}{Fast and sequence-adaptive whole-brain segmentation using parametric bayesian modeling}}.
\newblock {\emph{\JournalTitle{NeuroImage}}} \textbf{\bibinfo{volume}{143}}, \bibinfo{pages}{235--249} (\bibinfo{year}{2016}).

\bibitem{hsynex_data}
\bibinfo{author}{Rodrigues, L.} \emph{et~al.}
\newblock \bibinfo{title}{Synthetic ex vivo dataset of the human hypothalamus and its subregions}, \url{https://www.nitrc.org/doi/landing_page.php?doi=10.25790/bml0cm.161} (\bibinfo{year}{2023}).

\bibitem{taha2015metrics}
\bibinfo{author}{Taha, A.~A.} \& \bibinfo{author}{Hanbury, A.}
\newblock \bibinfo{journal}{\bibinfo{title}{Metrics for evaluating 3{D} medical image segmentation: analysis, selection, and tool}}.
\newblock {\emph{\JournalTitle{{BMC} medical imaging}}} \textbf{\bibinfo{volume}{15}}, \bibinfo{pages}{1--28} (\bibinfo{year}{2015}).

\bibitem{cicek20163d}
\bibinfo{author}{Özgün Çiçek}, \bibinfo{author}{Abdulkadir, A.}, \bibinfo{author}{Lienkamp, S.~S.}, \bibinfo{author}{Brox, T.} \& \bibinfo{author}{Ronneberger, O.}
\newblock \bibinfo{title}{3d u-net: Learning dense volumetric segmentation from sparse annotation} (\bibinfo{year}{2016}).
\newblock \eprint{1606.06650}.

\bibitem{IXI}
\bibinfo{title}{{IXI Dataset}}.
\newblock \bibinfo{howpublished}{\url{https://brain-development.org/ixi-dataset/}}.
\newblock \bibinfo{note}{Accessed: 2021-06-12}.

\bibitem{mueller2005alzheimer}
\bibinfo{author}{Mueller, S.~G.} \emph{et~al.}
\newblock \bibinfo{journal}{\bibinfo{title}{The alzheimer's disease neuroimaging initiative}}.
\newblock {\emph{\JournalTitle{Neuroimaging Clinics}}} \textbf{\bibinfo{volume}{15}}, \bibinfo{pages}{869--877} (\bibinfo{year}{2005}).

\bibitem{rodrigues2024hsynex}
\bibinfo{author}{Rodrigues, L.} \emph{et~al.}
\newblock \bibinfo{title}{H-synex: Using synthetic images and ultra-high resolution ex vivo mri for hypothalamus subregion segmentation} (\bibinfo{year}{2024}).
\newblock \eprint{2401.17104}.

\bibitem{iglesias2023easyreg}
\bibinfo{author}{Iglesias, J.~E.}
\newblock \bibinfo{journal}{\bibinfo{title}{Easyreg: A ready-to-use deep learning tool for symmetric affine and nonlinear brain mri registration}}.
\newblock {\emph{\JournalTitle{Scientific Reports}}}  (\bibinfo{year}{2023}).

\end{thebibliography}

\end{document}